\documentclass[namedreferences]{solarphysics}
\usepackage[hyperref,optionalrh,solaromanenum]{spr-sola-addons} 
\usepackage{graphicx}                    
\usepackage{color}                       
\usepackage{breakurl}                         
\usepackage[urlcolor=blue,breaklinks]{hyperref} 

\ifx \doiurl    \undefined \def \doiurl#1{\href{http://dx.doi.org/#1}{\textsf{DOI}}}\fi
\ifx \adsurl    \undefined \def \adsurl#1{\href{http://adsabs.harvard.edu/abs/#1}{\textsf{ADS}}}\fi
\ifx \arxivurl  \undefined \def \arxivurl#1{\href{http://arxiv.org/abs/#1}{\textsf{arXiv}}}\fi

\newcommand{\aap}{    {\it Astron. Astrophys.}}

\newcommand{\apj}{    {\it Astrophys. J.}}
\newcommand{\apjl}{   {\it Astrophys. J. Lett.}}

\newcommand{\grl}{    {\it Geophys. Res. Lett.}}

\newcommand{\nat}{    {\it Nature}}

\newcommand{\solphys}{{\it Solar Phys.}}

 \newcommand{\memsai}{    {\it Mem. S.A.It.}}
\chardef\us=`\_

\begin{document}


\begin{opening}

\title{Long-term Variations in the Intensity of Plages and Networks as 
Observed in Kodaikanal Ca-K Digitized Data}


\author[addressref={Muthu Priyal},corref,email={privee05@gmail.com}]{\inits{}\fnm{}\lnm{Muthu Priyal}}
\author[addressref={Jagdev Singh},corref,email={}]{\inits{Indian Institute of Astrophysics, Bangalore-560034}\fnm{}\lnm{Jagdev Singh}\orcid{}}
\author[addressref={Jagdev Singh},corref,email={}]{\inits{Indian Institute of Astrophysics, Bangalore-560034}\fnm{}\lnm{Ravindra B}\orcid{}}
\author[addressref={Rathiha S.K},corref,email={}]{\inits{Government Arts and Science college, Melur-625106}\fnm{}\lnm{Rathina S. K}\orcid{}}

%
\runningauthor{Priyal \textit{et~al.}}
\runningtitle{long-term variations}

\address[id={Muthu Priyal}]{Bharathiyar University, Coimbatore-641046, India}
\address[id={Jagdev Singh}]{Indian Institute of Astrophysics, Bangalore-560034, India}
\address[id={Rathiha S.K}]{Government Arts and Science college, Melur-625106, India}

\begin{abstract}
In our previous article (Priyal \textit{et al.}, {\textit Solar Phys.}, {\bf 289}, 127) we have discussed the details of observations and methodology adopted to analyze the Ca-K spectroheliograms obtained at Kodaikanal Observatory (KO) to study the variation of Ca-K plage areas, enhanced network (EN) and active network 
(AN) for the three solar cycles, namely 19, 20, and 21. Now, we have derived the areas of chromospheric features using KO Ca-K spectroheliograms to study the long term variations 
of solar cycles between 14 and 21. The comparison of the derived plage areas from the 
data obtained at KO  observatory for the period 1906 -- 1985 with that of MWO, NSO for the 
period 1965 -- 2002,  earlier measurements made by Tlatov, Pevtsov, and Singh 
(2009, {\textit Solar Phys.}, {\bf 255}, 239) for KO data and the SIDC 
sunspot numbers shows a good correlation. Uniformity of the data obtained with the 
same instrument remaining with the same specifications provided a unique opportunity to 
study long term intensity variations in plages and network regions. Therefore, we 
have investigated the variation of intensity contrast of these features with time at a  
temporal resolution of 6-months assuming the quiet background chromosphere 
remains unchanged during the period of 1906 -- 2005 and found that average intensity of 
AN, representing the changes in small scale activity over solar surface, varies with 
solar cycle being less during the minimum phase. In addition, the average intensity of 
plages and EN varies with a very long period having a maximum value during the Solar 
Cycle number 19 which is the  strongest solar cycle of 20th century.
\end{abstract}

%
\keywords{Sun, Chromosphere, Active and quiet, intensity variations, solar cycle}

\end{opening}

\section{Introduction}
     \label{sec:intro} 

The study of several types of long term activities on the surface of the Sun is very 
important to enable us to understand the internal dynamics of the Sun and the meridional flows 
\citep{choudhuri1995}. 
The meridional flows cause the observed systematic variation in the activity on the solar 
surface and recycling of the toroidal and poloidal components of magnetic field. The flow 
of material in the meridional plane direction, from the solar equator toward poles and from 
poles towards the equator deep inside the Sun, plays a significant role in the solar 
magnetic dynamo \citep{choudhuri1995, charbonneau2007}. The pattern of change in surface 
magnetic field, sunspots, H$\alpha$ filaments, Ca-K features and other solar activity indices, 
and its variation with time has implication for the study of meridional flows 
\citep{jiang2010, jin2012, sindhuja2014}.
	
\par The variations in solar irradiance might originate from number of reasons: i) change 
in the effective temperature of the Sun due to variations in the occurrence of 
faculae, plages and other active features \citep{gray1997}; ii) change in the size of the 
Sun \citep{ulrich1995}; iii) variation in activity with solar latitude changes with the solar 
cycle phase \citep{kuhn1990}. Various measurements of total solar irradiance (TSI) has 
indicated that it varies $\approx$0.2 \% over a solar cycle \citep{solanki2013}. But,  the Ca-K 
index representing chromosphere increases by $\approx$ 20 \% during the maximum phase of a 
solar cycle compared to the minimum phase \citep{white1981}. The large variations in 
Ca-K index compared to that in TSI makes it possible to study the periodic variations in 
the activity on the Sun from a large data set of chromospheric images obtained at various observatories. Using the measurements of solar spectral lines, \citet{gray1997} found that 
the variation in the effective temperature of the Sun is 1.5$\pm$0.2 K over a cycle. 
Further, \citet{penn2006} found that umbral magnetic field and temperature reduces during 
the minimum phase compared to maximum phase.

	Number of investigations about the long term variations in the chromospheric 
activity,  especially those pertain to variations in Ca-K plage areas were made using the 
Ca-K line images \citep{foukal1996, foukal2009, worden1998, tlatov2009, ermolli2009, priyal2014}. After identifying the plages in Mt. Wilson Ca-K images, \citet{foukal2006} found a linear relationship between the plage areas and the sunspot numbers. But, they did not find any 
good correlation between white-light facular areas and Ca-K plage areas. \citet{tlatov2009} compared the Ca-K data of the three observatories, namely Kodaikanal Observatory (KO), 
Mt. Wilson (MWO) and National Solar Observatory at Sacramento Peak (NSO/SP)  and found a 
high correlation between these data sets. Further, the comparison indicated that generally 
the plage areas determined from KO images are smaller by about 20 \% as compared 
those from MWO and NSO/SP data. This is probably due to different passband of the filter 
used or selected by the exit slit of the spectrograph.  \citet{foukal2009} 
compared the Ca-K indices derived from Kodaikanal, Mt. Wilson ad US National Solar Observatories 
and found that they show consistent behavior on the long time scales.
 After analyzing 1400 Ca-K line spectroheliograms taken during the period 
of 1980 -- 1996, \citet{worden1998} identified the plages, enhanced network(EN) and active 
network(AN) areas using the empirically determined values of threshold intensity for these features and found that plages and enhanced network areas occupy about 13 \% and 1 \% areas 
on the solar disk, respectively, during the active phase. In addition to this, the plages 
and enhanced network areas show strong rotation modulation of 27 days.

\citet{ermolli2009} compared various parameters such as, image quality, eccentricity, 
stray light, large scale inhomogeneities,  image contrast, etc., of Ca-K images taken from KO, 
NSO/SP, and MWO and concluded that it would be extremely useful to digitize the KO spectroheliograms  with a higher photometric accuracy and spatial resolution since the KO 
series is the most homogeneous and longest among these. It may be noted that KO had 
already started to digitize the Ca-K images with pixel resolution of 0.86 arcsec and 
16-bit digitization to achieve higher photometric accuracy \citep{priyal2014}. 

\citet{priyal2014} have analyzed the Ca-K line spectroheliograms for three solar cycles 
(19 -- 21) to study the variations of plage and network areas with solar activity on the Sun 
using the empirically determined values of threshold intensity contrast for the plages, 
enhanced network (EN) and active network (AN)  with respect to normalized quiet chromoshpere as done by \citet{worden1998}. They have normalized the intensity of the 
quiet chromosphere to 1 to measure the intensity contrast of various active features. 
This assumption is supported by the measurements of Ca-K index and Wilson--Bappu width 
of the line at the center of Sun which did not show change during the period of 
1974 -- 2006 \citep{livingston2007}. The details of the methodology to analyse the 
data and selected values of intensity contrast for different features are given in 
the earlier article by \citet{priyal2014}. 

It may be noted that most of the earlier studies consider the variation of plage areas 
with time and a couple of those studies investigate the variation of plage 
intensity with time. The methodology adopted by all the authors removes the effect of 
limb darkening only to identify the Ca-K line features. The procedure we have developed, 
yields the images free from all the effects such as limb darkening, vignetting due to 
instrument, stray light, and local defects due to photographic plate \citep{priyal2014}. 
Using the same procedure, we have extended the data analysis to the KO 
spectroheliograms for the period of 1906 -- 2005 to identify and determine the Ca-K plage 
areas, EN and AN areas on day-to-day basis.  The uniformity in the Ca-K  
spectroheliograms obtained at Kodaikanal over a century with the same instrument 
without any change in the optical components provides a unique dataset to study 
variation in the intensity contrast of plages and network with time. We therefore, 
have determined the average intensity contrast of these features with an interval 
of six months to study the variations, if any. 

In order to examine how much of contribution comes from the plages and network to the 
TSI, it is not only important to measure the plage and network element areas, it is 
equally important to measure the total or mean intensity of those regions on the 
daily basis. In this article, we first confirm the variation of the plage area with 
the solar cycle. Later, the computed mean intensity for the plages and networks is examined 
over ten solar cycles. In the end of this article we discuss the possible reason for 
the observed long term variations in the intensity of solar features.

\section{Data, Calibration, and Analysis}\label{sec:das}

The light feeding system is a 18-inch Focult siderostat which sends the sunlight to a 
12-inch, Cooke photo--visual objective. The objective lens makes the image of the Sun on the entrance slit of the spectroheliograph. 
The siderostat had a gravity based tracker in the 
past. It was replaced with motor during 90's. The rate of the gravity clock and frequency of 
the motor was adjusted to compensate the seasonal variations in the motions of the Sun in 
the sky. No guiding arrangement was introduced in the system at any time.

The spectroheliograph existing at the Kodaikanal observatory is similar to the one developed 
by Hale for the Yerkes observatory \citep{Hale1903}
The spectrograph has a set of two 4-inch prisms and two identical 6-feet focal length lenses 
to collimate the solar beam and focus the spectrum on to the exit slit. The width of the entrance 
and exit slits are maintained at 70 and 100 microns respectively, throughout the period 
of observations. The width of the exit slit permits 0.5~\AA~spectral pass-band centered around 
the Ca-K line allowing the violet and red emission wings of the line.  For a 
smooth movement to scan and build the image on a photographic plate or film, the whole of 
the spectroheliograph was made to move at a uniform speed across the Sun's image. This 
is achieved using a hydraulic and a pulley system on three rolling balls. The speed of 
this movement can be varied to adjust for the exposure time depending on the sky conditions 
and the time of the day. Generally, large exposure time is required in the morning hours 
compared to the noon.

By scanning over the Sun's image, the spectroheliogram was made 
in less than one-minute of time, generally around 30 seconds. Due to use of siderostat, the 
image rotates on its axis during the image build up on the photographic plate  
but the effect is negligible. It is around one-arcsec which is much less than the the 
seeing conditions at the Kodaikanal observatory. 
The images obtained have minor ellipticity 
(less than 1 \%) due to the drift of the image during the exposure. Any attempt to guide 
the image manually during the exposure time led to poor results. 

The full-disk spectroheliograms are made in Ca-K and H${\alpha}$ on daily basis.
In addition, by covering the image of the Sun by a round blackened plate (like coronagraph) 
at the focal plane of the solar image, chromospheric limb images are also made for the study 
of prominences. There are streaks in the Ca-K and H${\alpha}$ spectroheliograms.
The streaks in the spectroheliograms are due to the spectrograph slit. We 
have developed a software to compensate for the intensity vigneting (applied on digitized 
images) due to the instrument and a small shift of  light patch on the dispersing elements of 
the spectrograph. 

The orientation of the solar image is made after recording the image on the photographic plate. 
There is a mathematical formulation to determine the value of angle between north-pole of the 
Sun and the vertical top point of the image considering the place of observations, time and 
date of observations and p-angle. We had prepared tables for each day and time at an interval 
of six minutes. Interpolation was done if the observing time occurred between the two values 
of time. Then considering p-angle, date and time of observation, we identify the position 
of the North pole. North and south polarity was marked by placing the obtained image on 
the circular grid showing angle with a resolution of one degree. Thus the North--South poles 
were marked with a resolution better than one~degree. 

The detailed description of the Ca-K images obtained at the Kodaikanal Observatory, digitization and calibration procedures which involves finding center and radius of each image, centering of each image, image rotation, limb darkening correction using the background chromosphere without considering the active regions, normalizing the quiet chromosphere to uniform value of one, has been reported in our earlier article \citep{priyal2014}. 

\begin{figure}[!h]
\begin{center}
\includegraphics[width=0.45\textwidth]{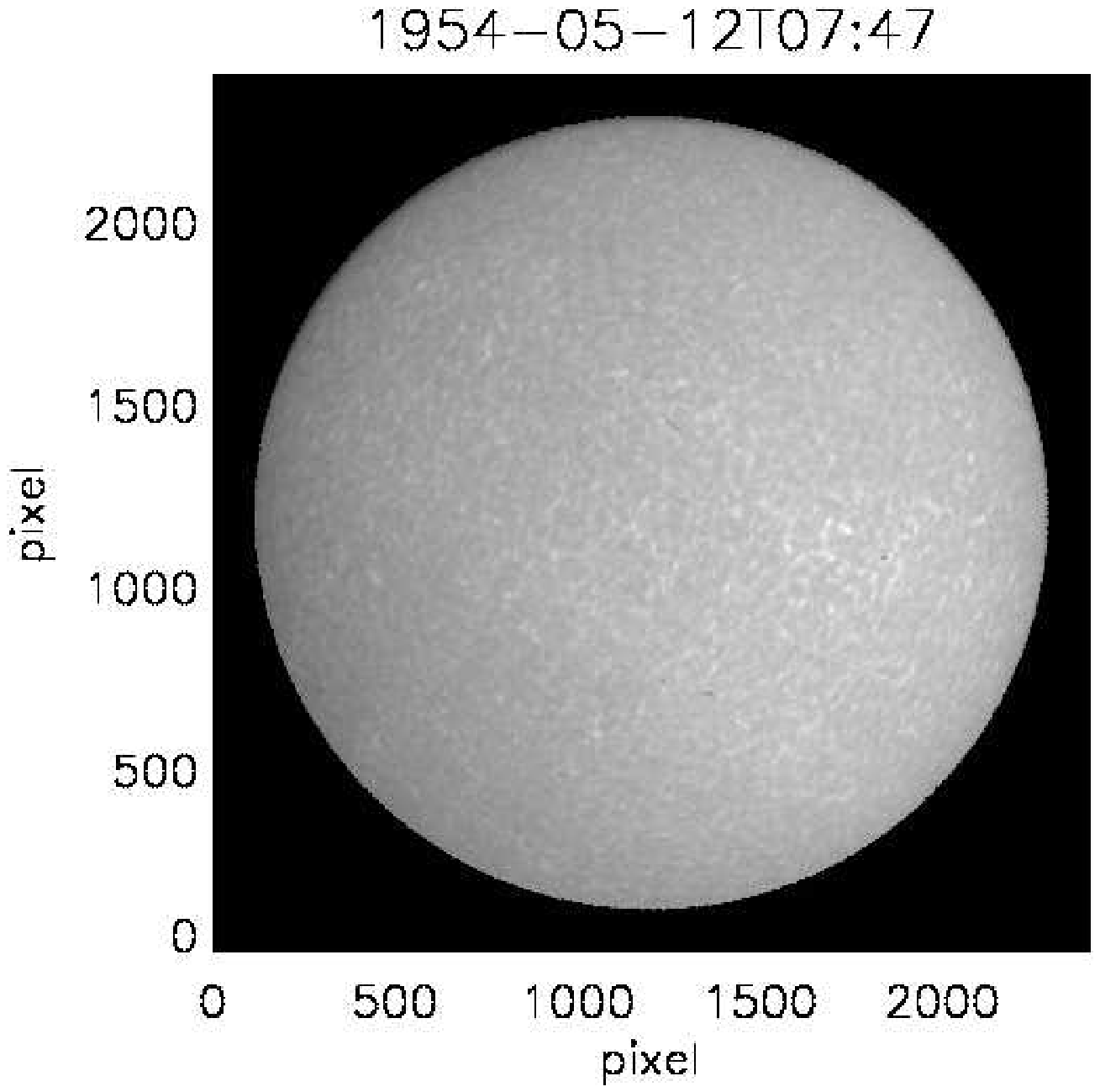}\includegraphics[width=0.45\textwidth]{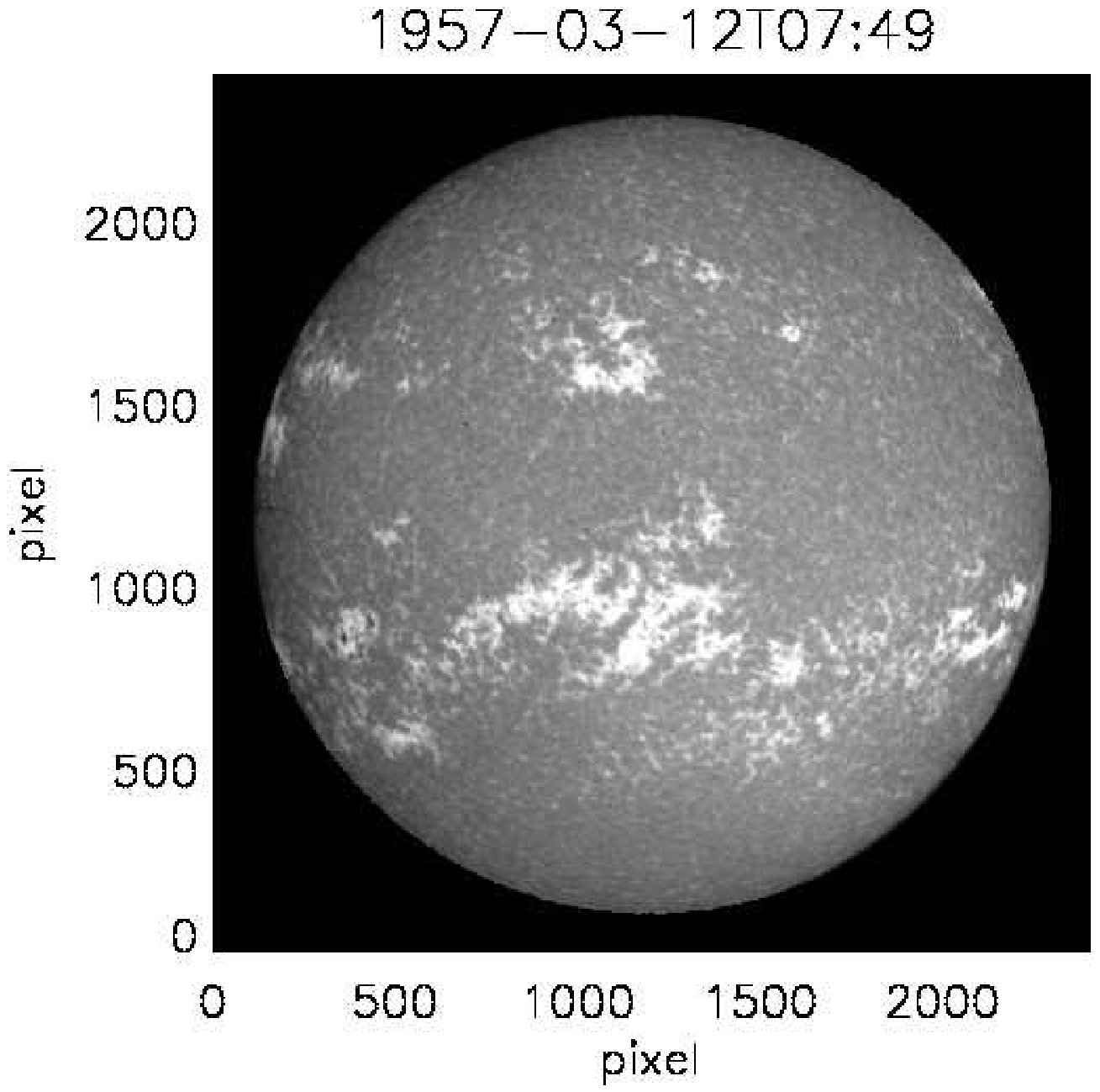} \\
\includegraphics[width=0.45\textwidth]{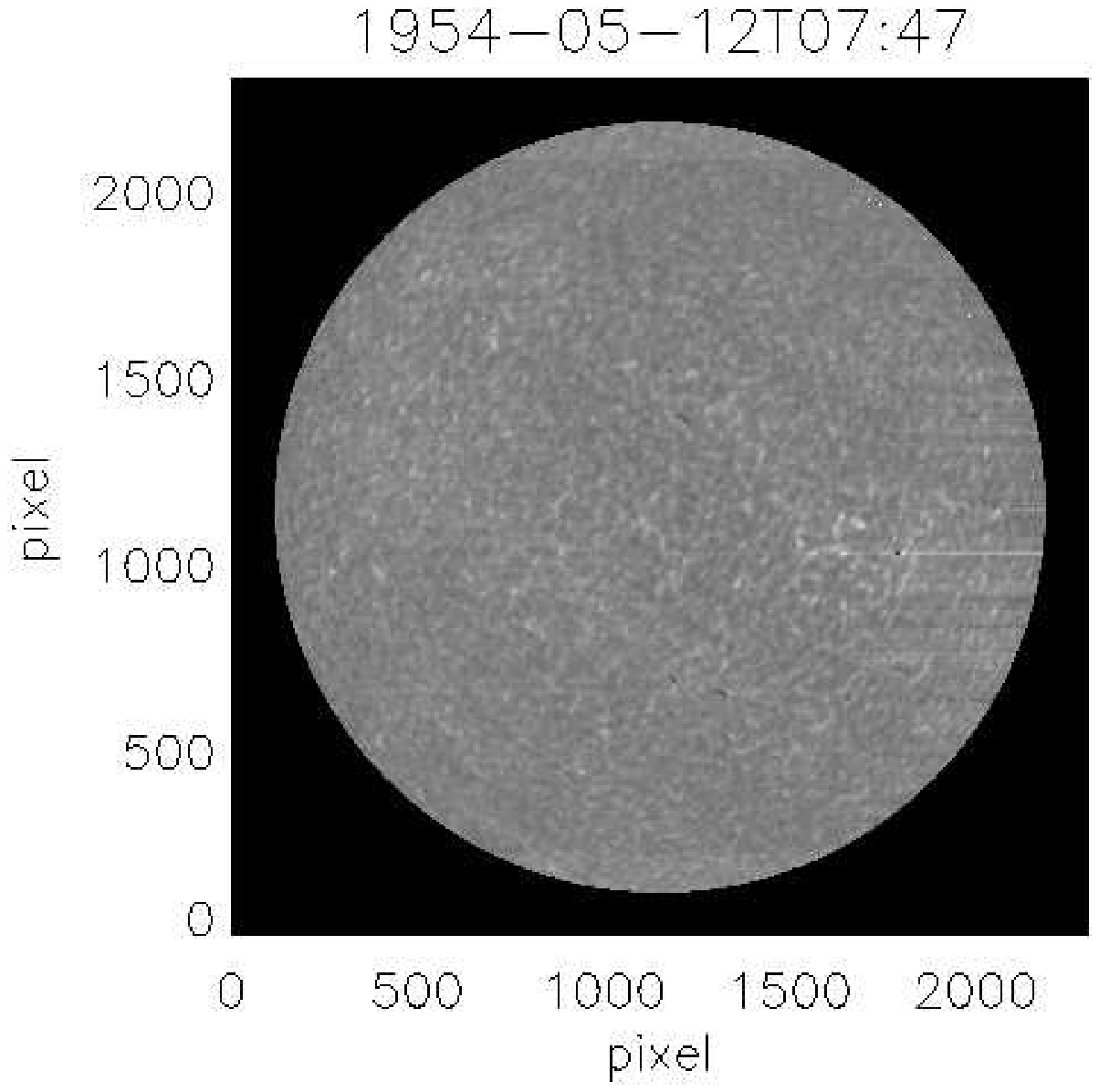}\includegraphics[width=0.45\textwidth]{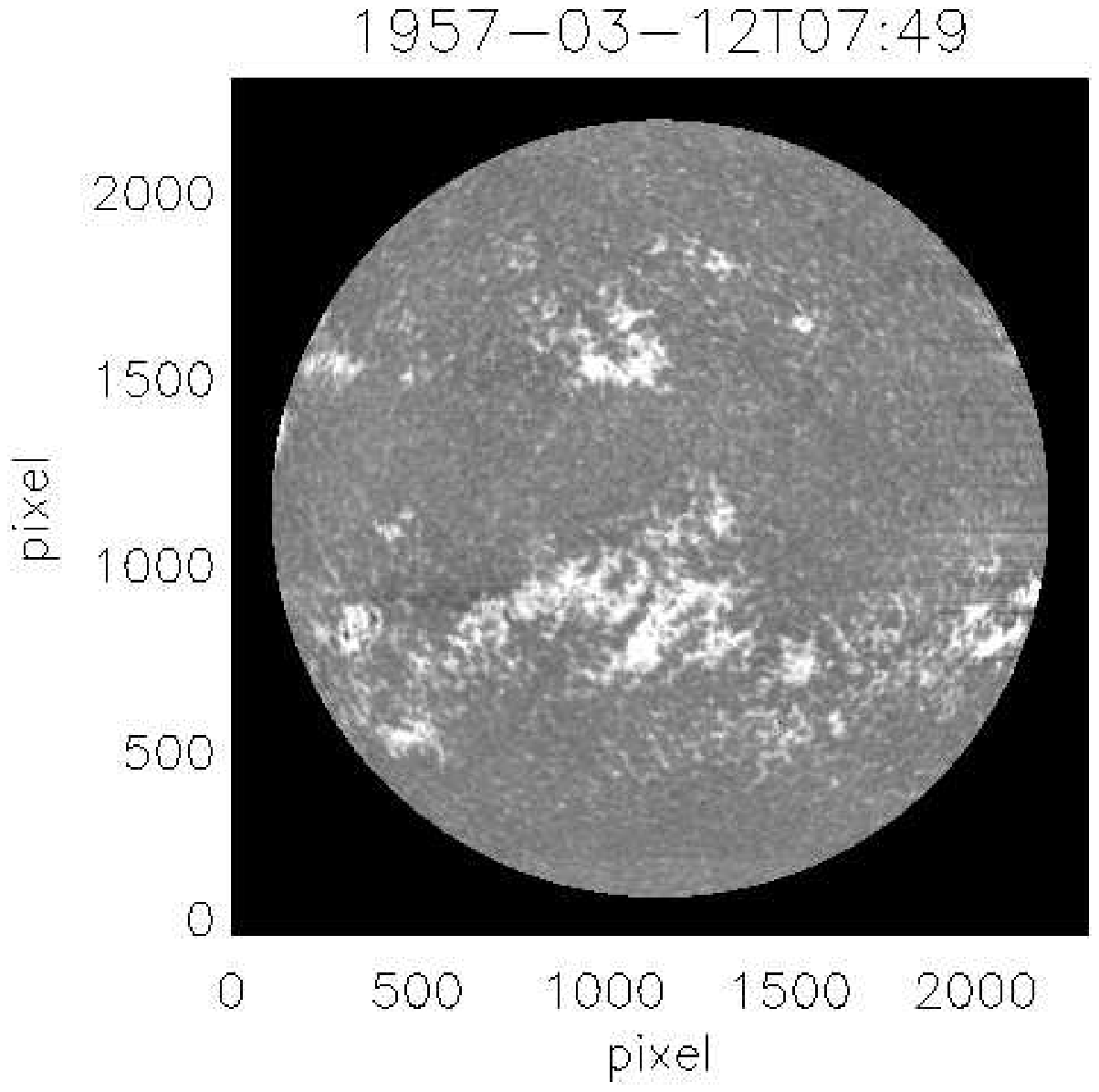} \\
\hspace*{1.0cm}
\includegraphics[width=0.45\textwidth]{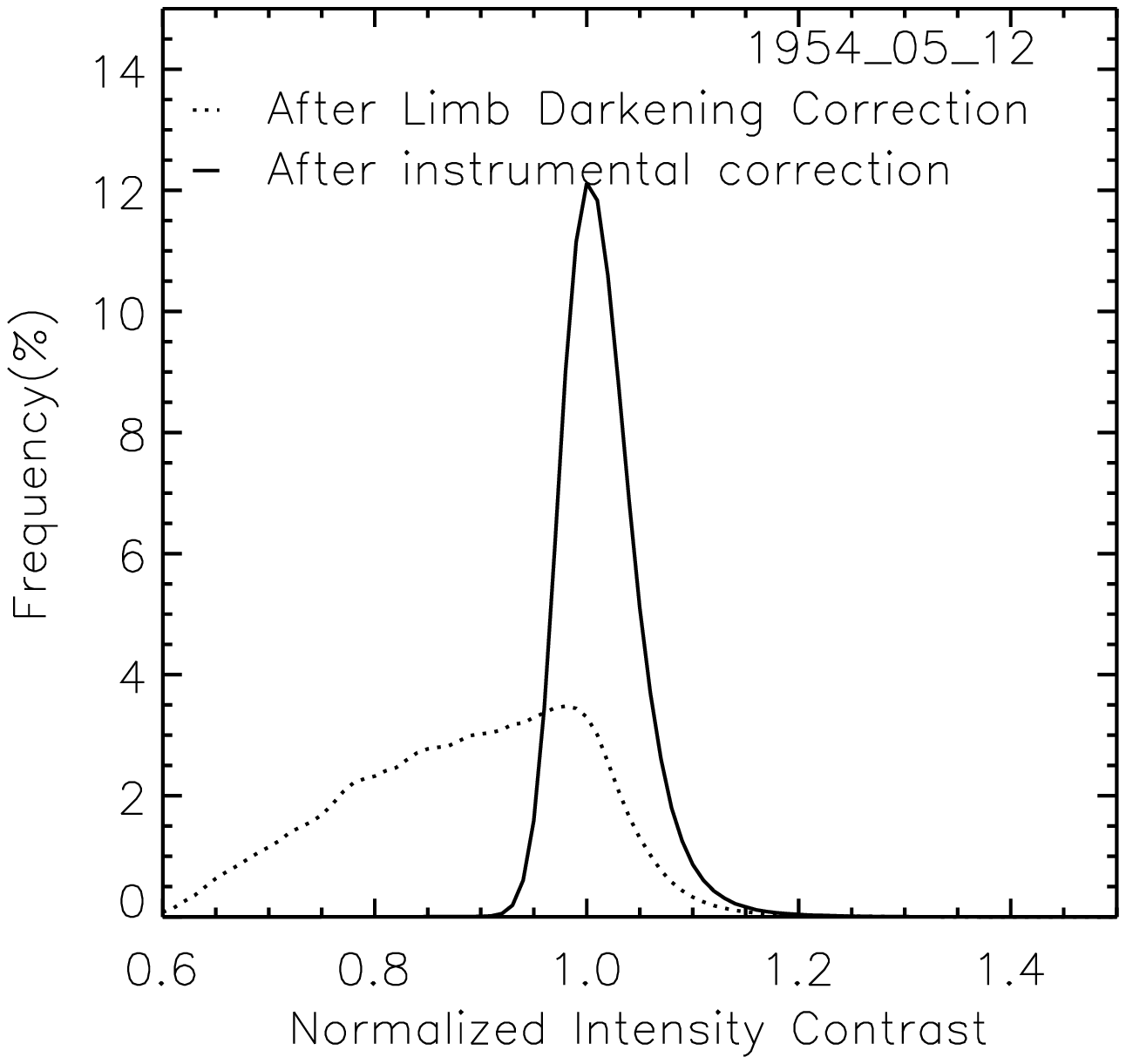}\includegraphics[width=0.45\textwidth]{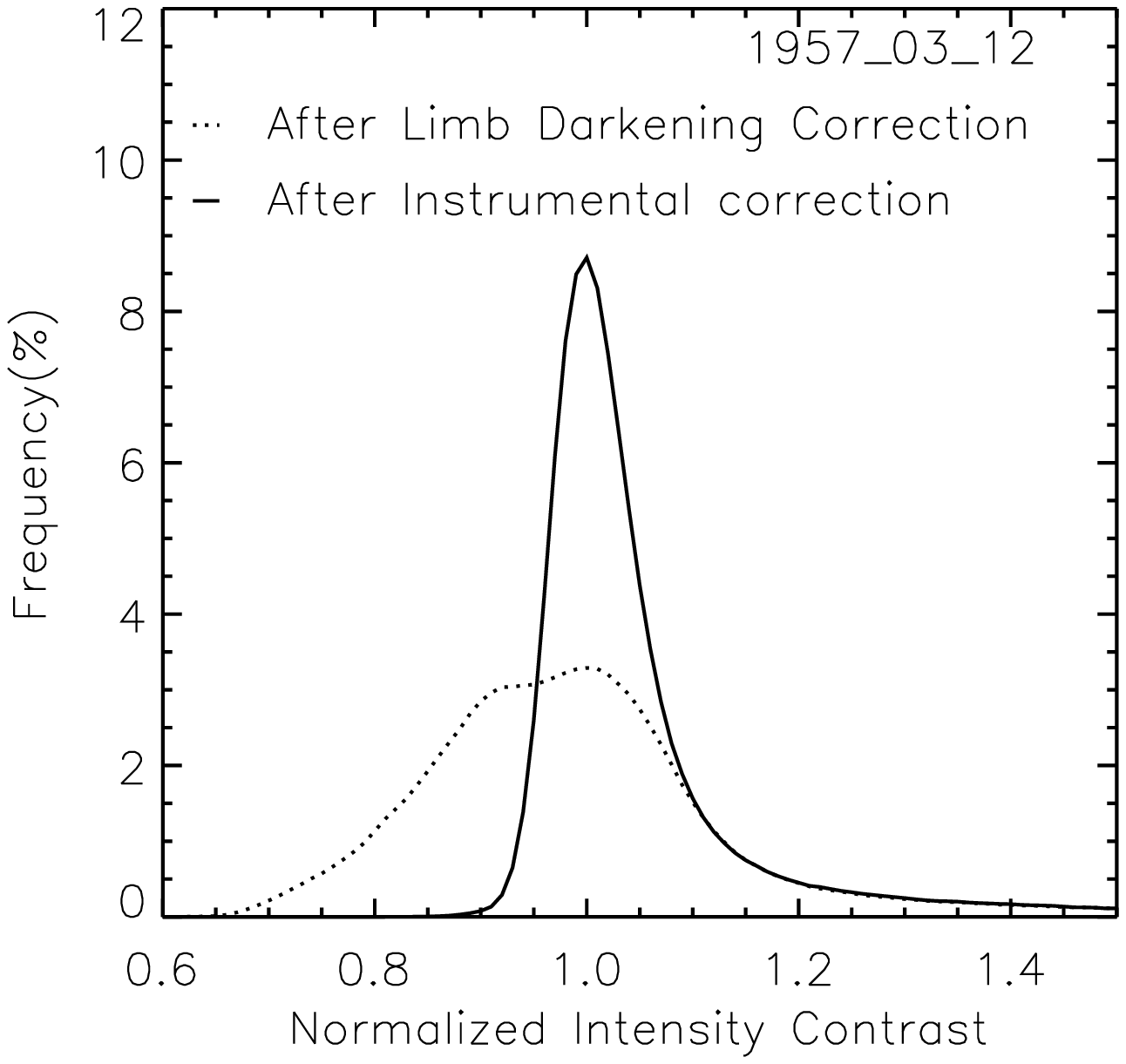} \\
\end{center}
\caption{Two panels in the top row show two typical images of the Sun in Ca-K line corrected 
only for the limb darkening effect taken during minimum (12 May 1954), and maximum  (12 March 1957) phase of the solar cycle. Two panels in the middle row show the same images after correcting for both limb darkening and intensity vignetting due to instrumental effects (displayed only up to 
0.95 R$_\odot$,). Panels in the bottom row show the frequency distributions of the intensity contrast for these images normalized to the quiet chromosphere.}
\label{fig:1}
\end{figure}

In addition to the procedure described in our earlier article, here we show the typical intensity contrast distribution of the images after removal of limb darkening and  instrumental effects during the minimum and maximum  phases of solar cycle in Figure~\ref{fig:1}. In the top row of the figure we show typical images after the removal of limb darkening for the minimum (1954) and maximum (1957) phases of the solar cycle. Visual inspection of the images indicate that on one side these appear brighter as compared to the other because of the residual effect of intensity vignetting in the images due to instrument as shown in our previous article \citep{priyal2014}. In the middle panel we show the same images after making correction for the limb darkening effect and intensity vignetting due to instrument as explained in the earlier article. These images show quiet chromosphere to be uniform over the whole image. Further, for comparison we show the intensity contrast distribution for these images in the bottom panel of this figure. The dashed line shows the intensity contrast distribution for the image corrected for the limb darkening and solid line indicates the same for the image corrected for both, limb darkening and instrumental effects. The final distribution of the intensity contrast is similar to the distribution obtained by \citet{bertello2010}. 

The bottom panel of this figure shows that the width of the intensity contrast distribution for limb darkening corrected images (dotted curve) is larger and irregular as compared to that of both limb darkening and instrumental effects corrected image (solid curve). Intensity vignetting in the image due to instrument makes the distribution broader and thus needs to be corrected.
All the measurements of the plages and networks have indicated that at any given time, during all phases of the solar cycle, the areas occupied by the quiet background chromosphere are more than the total areas occupied by the plages and networks including the quiet network on the solar surface. Therefore, it is expected that the peak of the intensity distribution curve of the image represents the background chromosphere. The intensity of the background chromosphere does vary with time as it represents general temperature of the Sun. Even if it varies with time, the variation in the background will be marginal and will not change the results of present study. We, therefore, after calibrating and correcting images for the instrument effects, have normalized the peak of the intensity distribution curve to one. The portion on the left side of the intensity distribution curve at the time of minimum phase of the Sun indicates that the values of intensity contrast vary from 0.9 to 1.0, representing the background chromosphere. Hence, the intensity contrast values between 1 $\pm$ 0.1 shows the scatter in the background chromoshpere. We have referred these areas as quiet background chromosphere.   The intensity contrast during the minimum phase generally varies from 0.9 to 1.25. The areas of intensity contrast values larger than 1.1 represent mostly the quiet and active network. The larger tail and area under the curve for larger values of intensity contrast during maximum phase of the solar cycle is due to the occurrence of more plages and networks during that period. 

One may note that the quiet-background chromosphere is normalized to unity for all the data so that variations in the intensity contrast of the plages and networks with time can be investigated.
The larger tail and area under the curve for larger values of intensity contrast during maximum phase of the solar cycle is due to the occurrence of more plages during that time. One may study variations of plage area and network area with time using intensity threshold values for different features,  intensity contrast distribution and area under the curve. But, we identified the features in each image using the intensity threshold and the area threshold values.

We developed software program to automatically identify the chromospheric features such as plages, enhanced network, active network and quiet network (QN) as has been done in 
\citet{worden1998} using the threshold value for the features and filling the areas within the intensity contours for the plages and enhanced networks.  Following \citet{worden1998}, we have determined empirically the values of threshold intensities for the plage and network areas for KO data. By using the method of  filling the intensity contours \citet{worden1998} found an average plage area to be larger by a factor of two than determined by others from MWO data. Hence we did not
fill the intensity contours but computed the total areas occupied by pixels with the defined values of threshold intensity. The plages are identified by using threshold values of intensity contrast determined empirically by using randomly selected images spread over the observing period and fixing threshold values of intensity contrast for each feature. We found that these threshold values were able to identify the features properly in about 90 \% of good images obtained during 1906 -- 2005. After visual examination of the analysed images, the remaining 10 \% of data where the features could not 
be identified properly were discarded. Further, we have determined the average intensity contrast 
of plages, EN, and AN on day to day basis. In the next section we show the variations in the areas 
and intensity of plages, EN, and AN on a long-term basis.

\section{Results}\label{sec:result}

We remind the reader that the chromospheric features with intensity contrast larger than 1.35 and adjacent combined area more than one arcmin$^{2}$ were considered as plages. Regions with combined areas less than one arcmin$^{2}$ and larger than 0.3 arcmin$^{2}$ and intensity contrast larger than 1.35 are considered as EN elements. The regions with combined area 
less than 0.3 arcmin$^{2}$ and intensity contrast larger than 1.35 were considered AN elements along with regions of intensity contrast between 1.25 and 1.35. Finally, areas with intensity contrast between 1.15 and 1.25 are taken as the quiet network (QN) elements. By normalizing the intensity of quiet chromosphere to unity value for whole of the data we have determined the average intensity of the plages and network with respect to normalized background, quiet chromosphere. In the following we discuss the long term variations in their areas and intensity contrast.

\subsection{Variations of Ca-K plage Areas and its Comparison with Other Data}

The temporal variations of plage area is well studied in the past using MWO data 
\textsf{ftp.ngdc.noaa.gov/STP/SOLAR$\_$DATA\newline/SOLAR$\_$CALCIUM/DATA/Mt$\_$Wilson/}
and we compared our detected plage area with that of MWO and the SIDC sunspot number. 
We determined the plage area in units of fraction of the solar disk on day-to-day basis 
for the period 1906 -- 2005. The data for the period of 1986 -- 2005 has large gaps due to sky conditions and large number of defective photographic plates. 

\begin{figure}[!h] 
\begin{center}
\includegraphics[width=\textwidth]{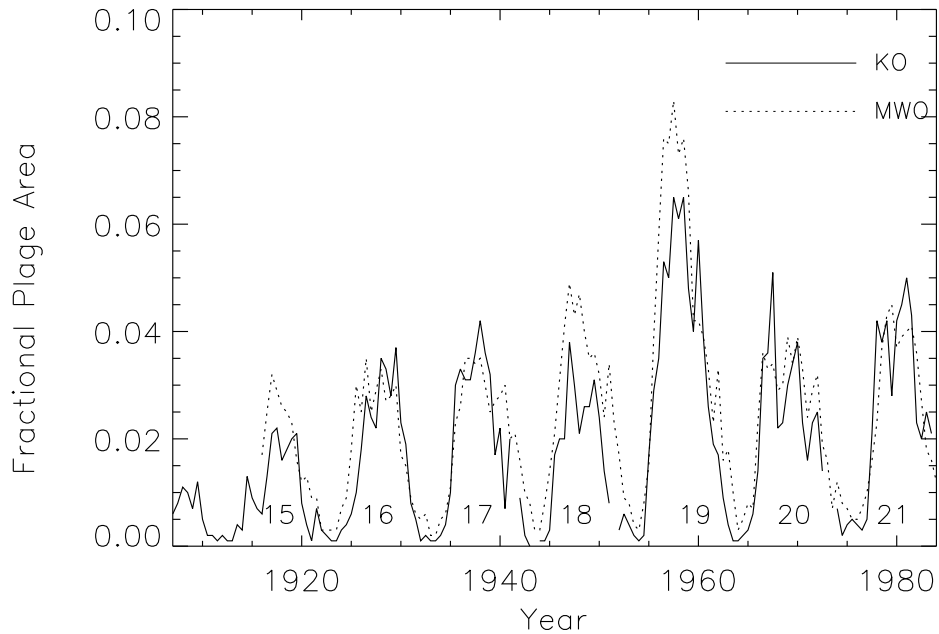}
\end{center}
\caption{The half yearly averaged Ca-K plage area is shown for eight solar cycles. The solid line
represents the KO and the dotted line represents the MWO Ca-K data.}
\label{fig:2}
\end{figure}

To compare the measured KO plage area with other data, we have computed the average plage 
areas on half yearly basis. Figure~\ref{fig:2} show half yearly averaged plage areas (solid line)
for KO data and the corresponding MWO data (dotted line) in the plot. Half yearly averaged plage 
areas indicate that there is general agreement between the KO and MWO data but about 15 -- 20 \% smaller plage areas for the KO data as compared to MWO data for the two Solar Cycles 18 and 19 during the maximum phase. \citet{tlatov2009} also find a similar behavior for the Solar Cycle 19 and interpret this type of abnormal behavior of Cycle 19 due to masking effect of sunspots in calculation of plage indices.

\begin{figure}[!h] 
\begin{center}
\includegraphics[width=\textwidth]{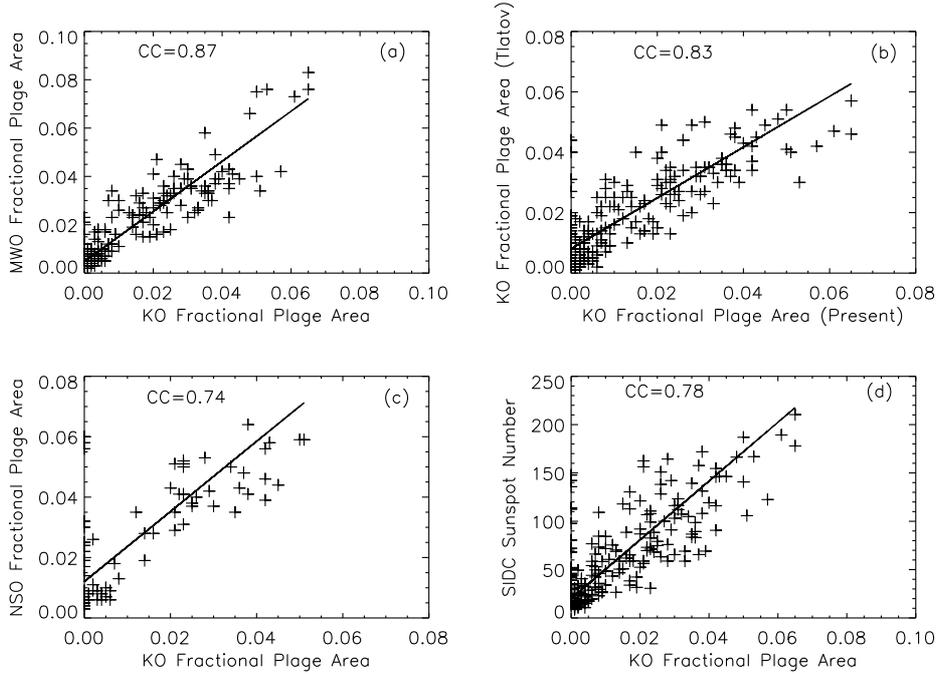} 
\end{center}
\caption{Panel a of this figure shows half-yearly averaged plage areas of MWO data is plotted against KO plage area determined using the high-resolution images for the period of 1915 -- 1985. Panel b shows a plot of TPS values against PSR values for the KO data where TPS is for both plages and enhanced network determined by Tlatov et al. (2009) and PSR is for plages derived by us.  Panel c shows the plot of NSO index against PSR values for the period of 1965 -- 2002. In panel d we show a scatter plot between the half-yearly averaged sunspot numbers and the corresponding plage area along with a linear fit to the data.}
\label{fig:3}
\end{figure}

In Figure~\ref{fig:3} we show the scatter plot between the half yearly averaged plage areas computed by us (here after called PSR) and MWO Ca-K index, KO Ca-K index (TPS, \citet{tlatov2009}) and NSO Ca-K index. Panel a shows half-yearly averaged plage areas (KO plage-index) determined using the high-resolution images versus the MWO Ca-K index for the period of 1915 -- 1985; panel b indicates the PSR versus TPS values for the KO data. Here PSR is for plage area derived by us and TPS includes both plages and enhanced network determined by Tlatov et al. (2009).  Panel c shows the PSR values against the NSO index for the period of 1965 -- 2002. Further, in panel d we show a scatter plot between the half yearly averaged sunspot number and the corresponding plage area along with a linear fit to the data. All the four linear fits to the respective data sets show good correlation between different measurements and correlation coefficient greater than 0.75 with confidence level more than  
99 \% in all cases. The aim is to show the good correlation between different measurements and to prove the reliability of the methodology adopted and further investigate the variations in the networks and intensity contrast of plages and networks with time. A linear fit seen in panel a to the data gave the following relation between the MWO and KO results.

\begin{equation}
MWO\_PA = 1.039(KO\_PA) + 0.004 
\end{equation}

Where MWO\_PA and KO\_PA are the plage areas from Mount Wilson Observatory and Kodaikanal Observatory data, respectively. The relation between these two measurements indicate that on an average plage area values from KO data are $\approx$10 \% less than those from MWO data, compared to 20 \% difference in earlier measurements by \citet{tlatov2009}. The decrease in difference between KO and MWO values may be because the present measurements are using data digitized with 
high-resolution compared to the data  digitized with low-resolution used by \citet{tlatov2009}. 

\citet{foukal1996} find the fractional plage areas of the visible Sun were about 8 \%, and 4 \%, during the maximum phase of Solar Cycle numbers 19 and 21, respectively. \citet{tlatov2009} also find that maximum fractional plage area including the enhanced network during the period of 1957 -- 1958 to be about 8 \%.  In contrast, \citet{worden1998} found that the plage and enhanced network typically cover about 13 \%, and 10 \%, respectively, of the solar disk during the maximum phase of Cycle 21 and 22, moderate cycles as compared to the Cycle number 19. Further, they reported that during moderate and minimum activity levels, the total plage and enhanced-network areas can reach zero, but the active network can still cover a large portion of the solar disk. Thus there are large variations in the fractional plage area determined by various techniques adopted. It may be noted that there is no absolute way of defining the plages or networks by assigning the threshold values of intensity contrast. These will vary for the data of different observatories depending upon the specifications of the instrument and passband used for observations. It is very important that all the data used in the analysis should be 
taken with the same instrument and analysed uniformly to study the temporal variations as in the present study. Earlier authors have also fixed the threshold values of intensity contrast empirically as we have done now by examining  number of images of the Sun taken at different times. We find that half yearly averaged maximum fractional plage area for a given cycle appears to be related to the strength of the solar cycle represented by the sunspot numbers and it varies from 4 to 7 \%, during the period 1906 -- 2005.

\subsection{Butterfly Diagram for Plages}

\begin{figure}
\begin{center}
\includegraphics[width=\textwidth]{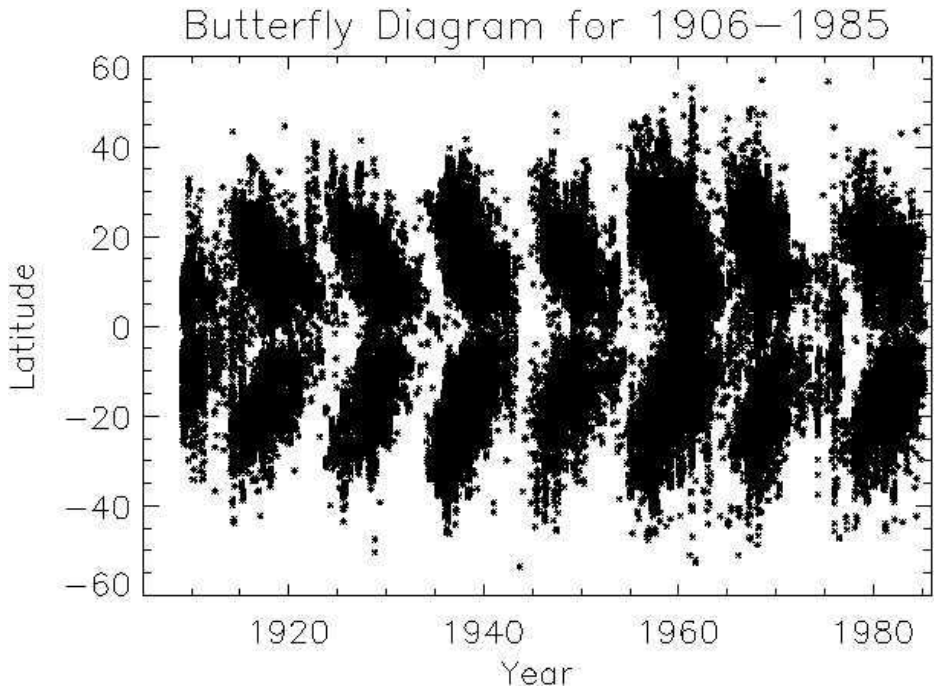} 
\end{center}
\caption{Butterfly diagram for the Ca-K plages is plotted for plages of all the sizes. The 
centroid latitude values of each plage was taken for the plot.} 
\label{fig:4}
\end{figure}

\begin{figure}
\begin{center}
\includegraphics[width=\textwidth]{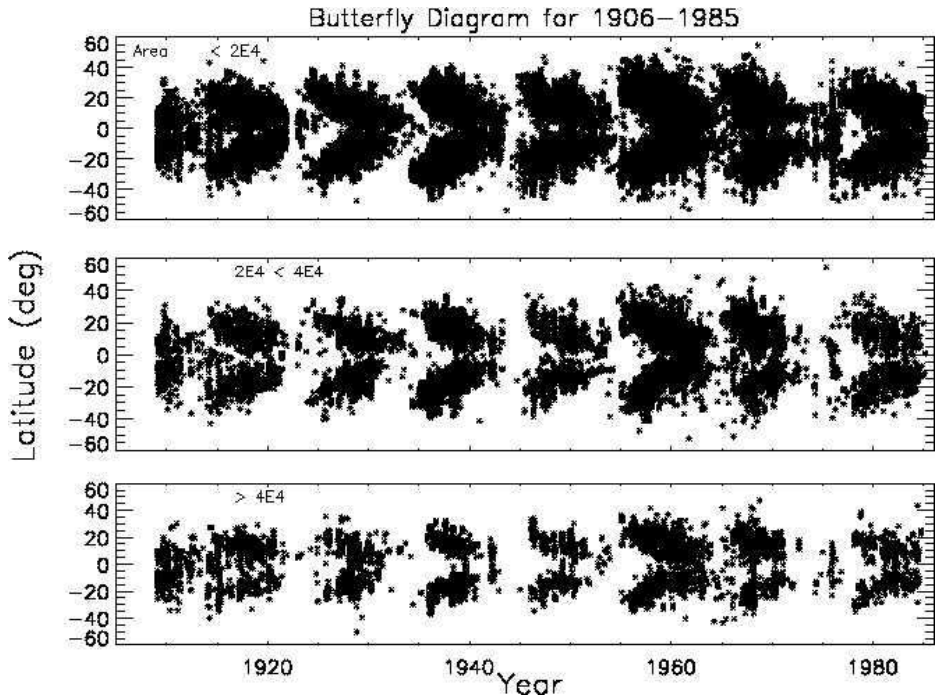}
\end{center} 
\caption{Butterfly diagram after grouping the Ca-K plages in three sets. Top panel of the figure shows the occurrence of plages with area $<$ 2$\times$10 $^{4}$ pixels equal to 4.1 arcmin$^{2}$, middle panel is for the plages with areas between 4.1,--8.2 arcmin$^{2}$ and the bottom panel shows the location of plages with areas  $>$ 8.2arcmin$^{2}$ for the period 1906,--1985.}
\label{fig:5}
\end{figure}

To make the butterfly diagram we need to know the latitude and area of each plage 
on day to day basis. After the identification of pixels with intensity contrast 
larger than 1.35 corresponding to plage regions, we converted the image into 
binary format i.e., plage pixels as 1 and remaining pixels as zero. By using 
IDL's \textsf{LABEL$\_$REGION.PRO}, we found groups of pixels whose value is one, 
in other words without any gap, each group representing a plage region. In every 
region defined as plage, the number of pixels were counted to compute the area 
of the plage and the regions with area less than one~arcmin$^{2}$ were not
considered as plages but considered as EN. Then operating our code for each 
identified plage region, and by using center of mass concept, we computed the 
centroid of each plage region in terms of pixel coordinates and also its area. 
Then we converted the pixel coordinates to heliographic coordinates (latitude and 
longitude) considering the apparent size of image on that day. Following this 
procedure, we determined the centroid of all the plages for all the data. In 
Figures~\ref{fig:4} and \ref{fig:5} we plot the latitude centroid of each plage as a 
function of time for the period of 1906 -- 1985, popularly known as butterfly diagram. 
Figure~\ref{fig:4} is considering all the plages of different sizes whereas in 
Figure~\ref{fig:5} we have grouped the plages in three sets. Visual inspection of 
both the figures indicate that all size of plages occur in large numbers during the 
stronger solar cycles as compared to the weaker cycles. The Solar Cycle number 19, 
the strongest Cycle of 20th century has more number of plages with larger areas and occur at 
larger latitude belt.

\begin{figure}
\begin{center} 
\includegraphics[width=\textwidth]{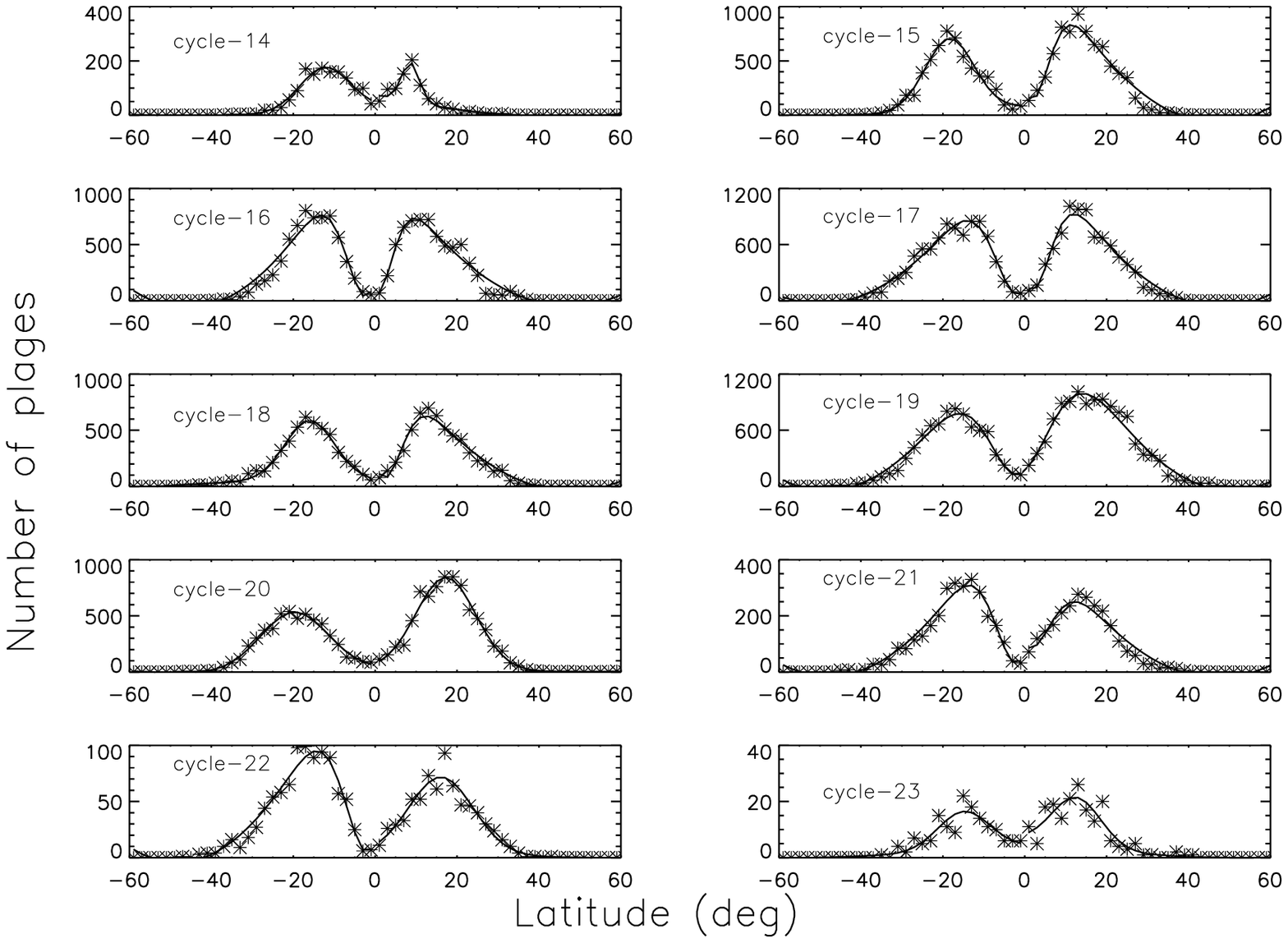} 
\end{center}
\caption{Number of plages plotted as a function of latitude for each Solar Cycle (14,--23) during the period of 1906,--2005.}
\label{fig:6}
\end{figure}

\begin{figure} 
\begin{center}
\includegraphics[width=\textwidth]{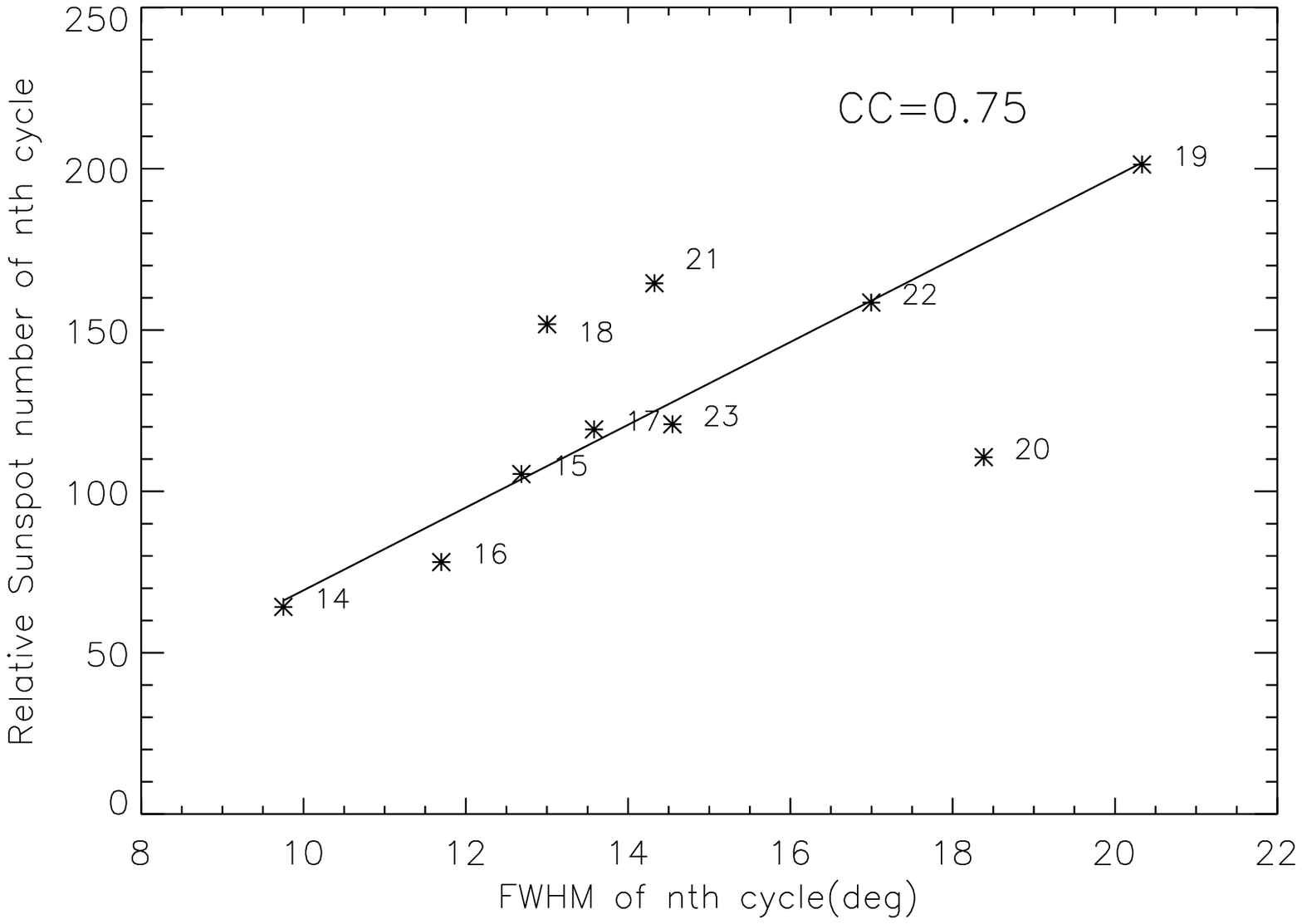} 
\end{center}
\caption{A plot of the maximum relative sunspot number in each cycle against FWHM of the distribution of plages in degrees indicating the spread of plages along the latitude of the Sun for the corresponding solar cycle.}
\label{fig:7}
\end{figure}

\begin{figure}
\begin{center} 
\includegraphics[width=0.5\textwidth]{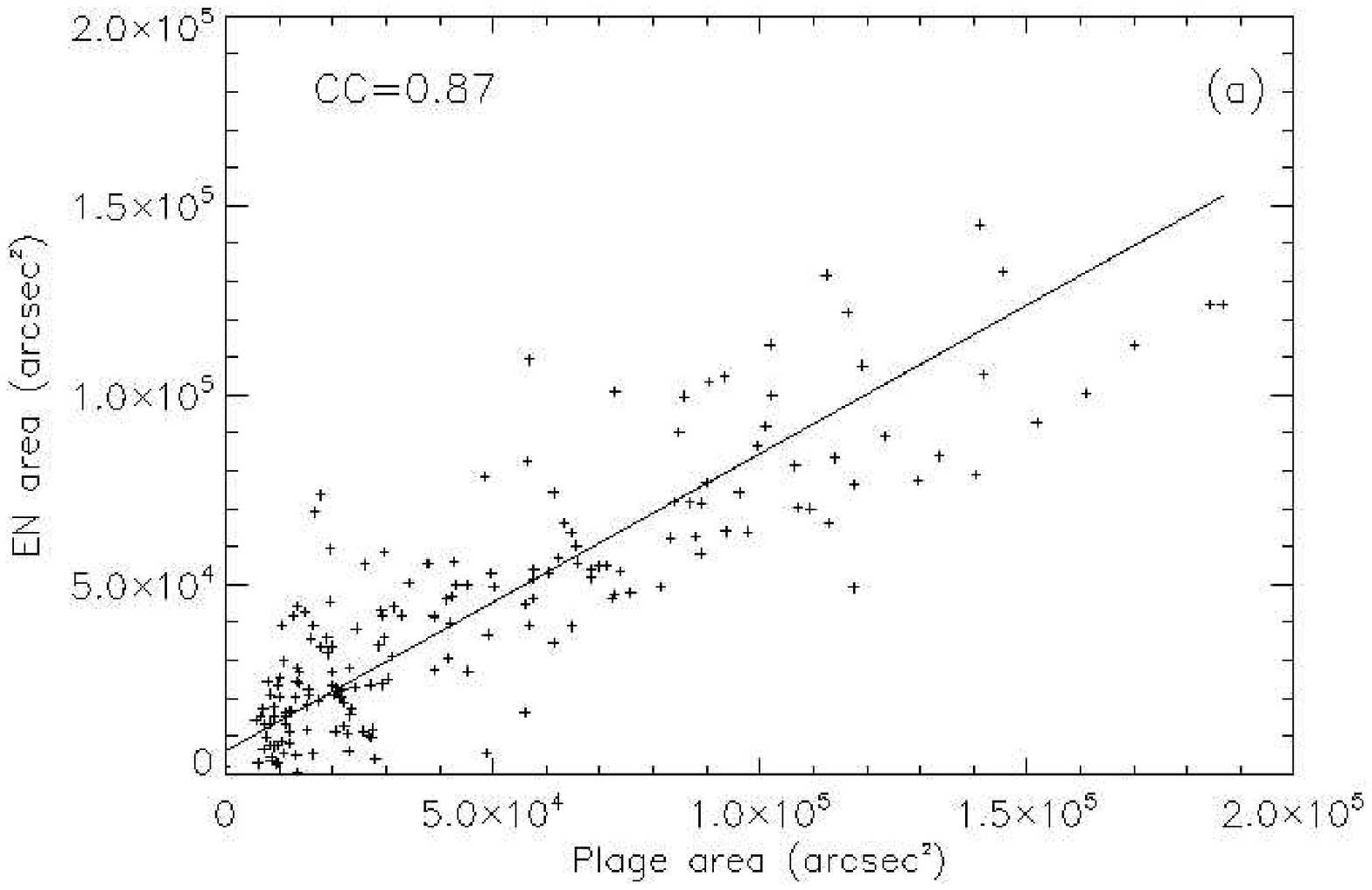}\includegraphics[width=0.5\textwidth]{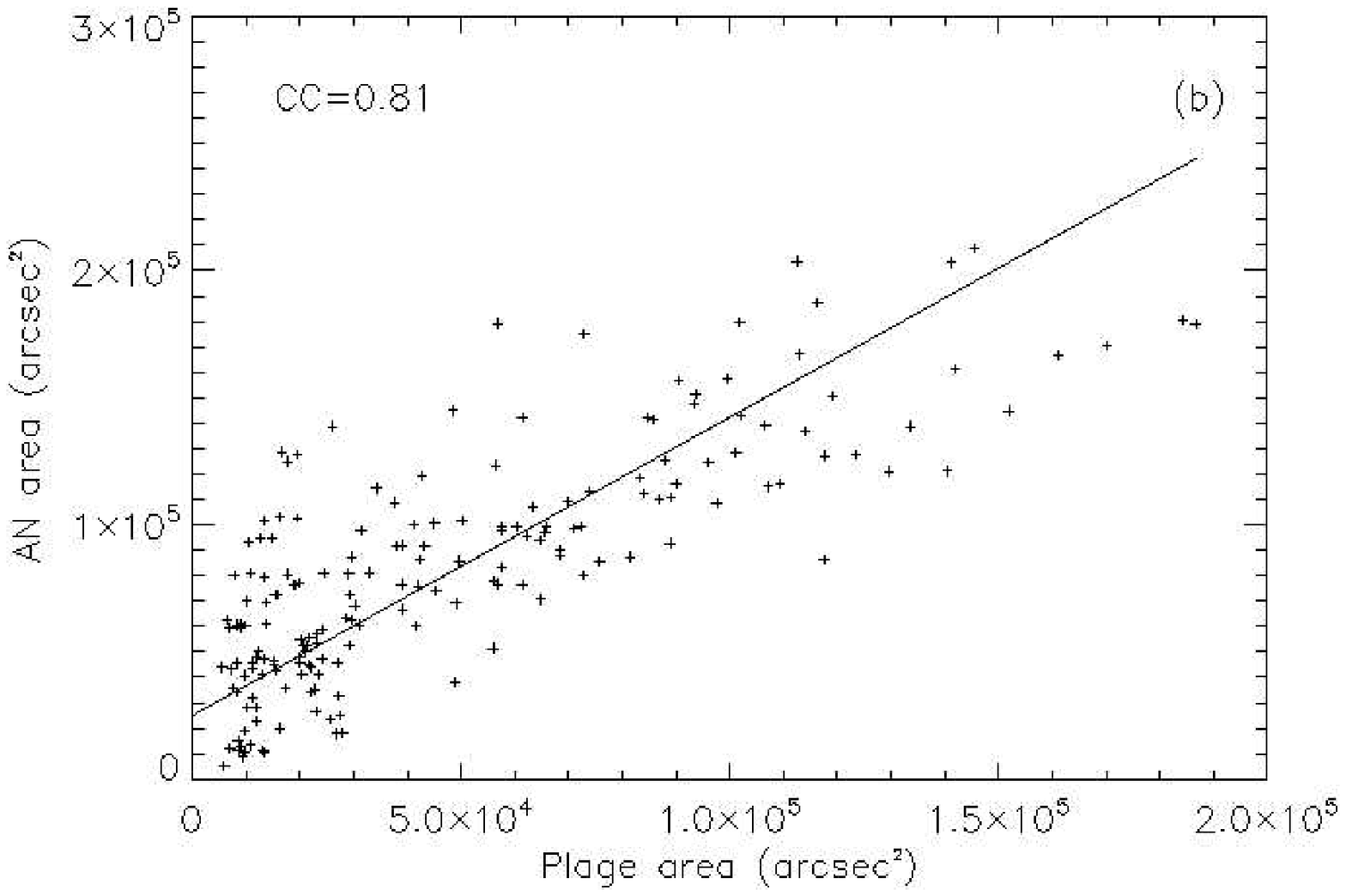} 
\end{center}
\caption{Left panel shows a plot of half yearly averaged area of enhanced network versus that 
of plages. Right panel shows the same for active network against plages.}
\label{fig:8}
\end{figure}

To study quantitatively the coverage of plages along latitudes, we have plotted the number of plages as a function of latitude for each solar cycle (Figure~\ref{fig:6}) and for both the hemispheres separately. The FWHM of Gaussian fit to the frequency distribution  for the northern and southern hemispheres for each cycle represents the spread in latitude. The plot of FWHM averaged over both the hemispheres for each cycle against maximum sunspot number for the corresponding cycle shown in Figure~\ref{fig:7}  indicates a correlation between these two. A linear fit to the data gave a correlation coefficient value of 0.75 indicating a confidence level of 99 \%. The linear fit indicates that FWHM of the spread along latitude is $\approx$20.5$^{\circ}$ for 19th solar cycle, the strongest cycle of 20th century and 
$\approx$10$^{\circ}$ for Cycle 14, the weakest cycle of this period. This implies that plages occur over larger latitude belts for stronger solar cycles as 
compared to that for weak cycles.  Further, we show the scatter plots between plage areas and EN as well as plage areas and AN networks  
(Figure \ref{fig:8}). We observed a linear relationship between the plage areas and 
EN, AN areas. 

\subsection{Long-term Variations in Intensity of Plage and Network Regions}
We have made an attempt to study the long term variations in the intensity of plages and network regions with respect to normalized intensity of quiet chromosphere. While carrying out this exercise we assumed that the quiet chromosphere do not change with the phase of solar cycle. 
Observations made by \citet{livingston2007} over a period of 30~years show that the Ca-K index measured at disk center does not vary significantly with the cycle of solar activity.
It may be noted that intensity contrast for plages and networks are different for different instruments and passband of the filter or selected wavelength band for spectroheliograms and thus any change made in the experimental set up is likely to affect the uniformity of the data.  
During the 100-year period covered by the Kodaikanal observations, no substantial changes were 
made to the instrumentation that may have altered the overall quality of the data. As stated earlier we first identified different chromospheric features using different intensity threshold values as defined in Section 3 of this article. We computed the daily average intensity contrast of plage by considering all the pixels in the plage region. Similarly, we computed the average intensity contrast of the EN and AN for each day of the data. Due to the dependence of plage and network intensity on the underlying magnetic fields \citep{ortiz2005, sivaraman1982}, the variation in the intensity of plages and network with long periods is likely because of the changes in the solar dynamo processes. Further, the intensity of plage and network is likely to vary on daily basis due to growth and decay of these features. 

\begin{figure} 
\begin{center}
\includegraphics[width=\textwidth]{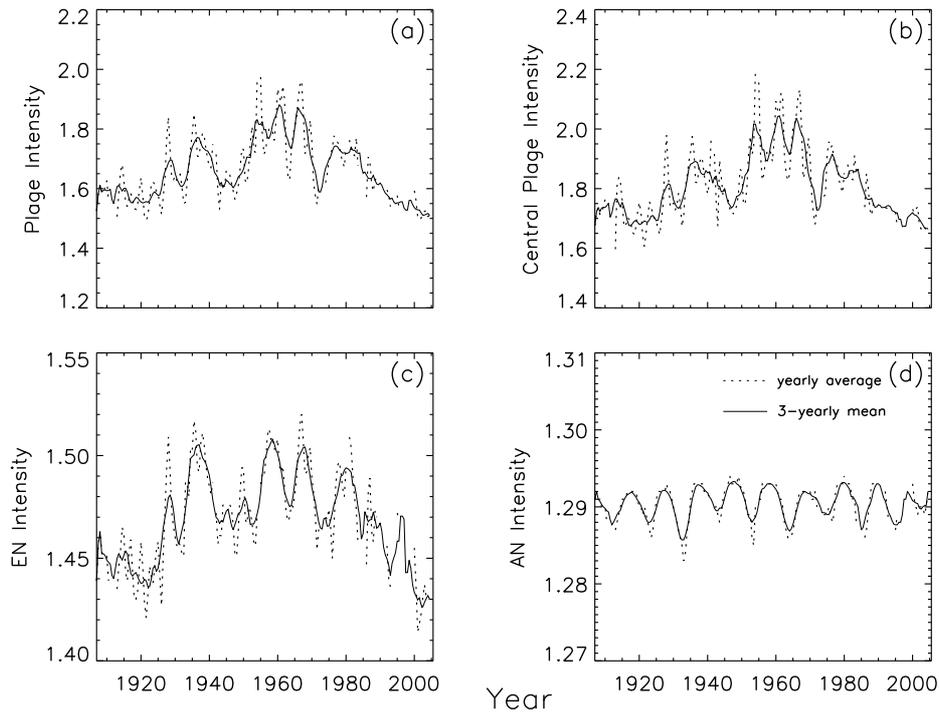}
\end{center}
\caption{ Panel a shows half yearly averaged KO plage intensity for the period 1906 -- 2005. Panel b shows the variations in the averaged intensity contrast of the central region of plages defined by intensity contrast $>$1.5. Panels c and d show average intensity contrast variation for the EN and AN networks, respectively.} 
\label{fig:9} 
\end{figure}

To study the long term variations in the intensity of plages and network, we
have computed average value of the intensity on half yearly basis for the period of 1906,--2005. We show the averaged intensity of plages, EN and AN as a function of time in Figure~\ref{fig:9}. We have also plotted the three year average of the intensity contrast to smoothen the data to view solar 
cycle variations.

Panel a in Figure~\ref{fig:9} shows the intensity of the plages appears to vary with the solar cycle phase and with some other quasi-periods. It may be noted that intensity contrast of plages with time definitely indicates that it varies with very long period, probably on time scale almost equal to the length of observations or more. The average intensity contrast of plages vary between 1.55 to 1.90 with some pattern during the period 1906 -- 2005. We have also studied the long term variations in the intensity of central region of plages considering the area represented by intensity contrast greater than 1.5 instead of intensity contrast of larger than 1.35 for plages. Yearly averaged intensity contrast of central region of plages at an interval of six months seen in panel b of Figure \ref{fig:9} indicates that average intensity varies between 1.7 to 2.1 
during this period. Panel c of this figure shows that the intensity of EN varies with the similar pattern as that of plages but with less amplitude. Panel d indicates that the average intensity of AN, which represents small-scale activity over the whole of the solar surface, varies with very regular pattern slightly different than that of plages and EN. Generally, the intensity of AN is more during the active phase as compared to that during the quiet phase.
The intensity of AN varied by $\approx$1 \%, from minimum value 1.286 (yearly averaged) to the maximum 1.293 (yearly averaged) during the period 1906 -- 2005.

\par To determine the quasi-periodicities we have computed the power spectral analysis of the time series by interpolating the few missing data points. At the first stage itself the time series is subtracted by the mean value.  Then the computed power spectra was normalized. Statistically significant peaks in the power spectrum were found by assuming a Gaussian distribution for each point in the time series. Every point in
the power spectrum will have a $\chi^{2}$ distribution with two degrees of freedom (DF). The significance level (sig) in the power spectrum was computed as \citep{torrence1998},

\begin{equation}
Pow_\mathrm{sig}=\frac{\sigma^{2}\chi^{2}(1-sig, DF)}{\frac{N}{\mathrm{2}} DF}
\end{equation} 

\begin{figure} 
\begin{center}
\includegraphics[width=\textwidth]{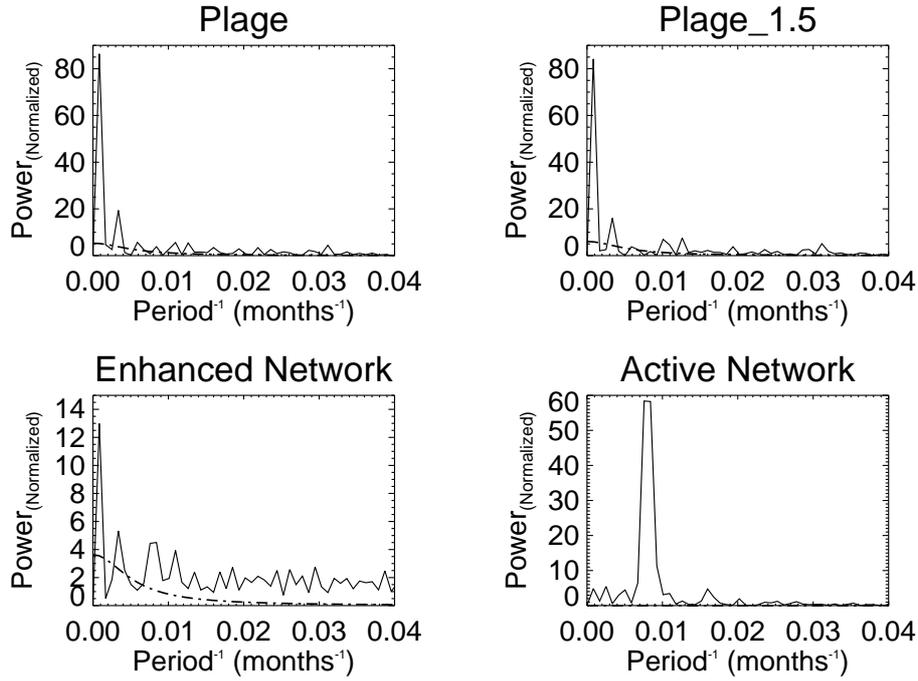} 
\end{center}
\caption{Figures show the power spectra of time series of averaged intensity contrast for plages, central region of plages, EN and AN intensities along with 99.9 \%, confidence level curve in each panel, respectively.}
\label{fig:10}
\end{figure}

Where N is the number of data points.
In Figure~\ref{fig:10} we show the power spectra for the time series of plage, central region of plage, EN and AN intensities along with 99.9 \%, confidence level. The average intensity of plages, central region of plages, and EN indicate a periodic behavior of almost equal to the period of observations. The peak at $\approx$90 years is very prominent. The average intensity profiles of plages, central region of plages, and EN also exhibits the same behaviour of $\approx$90 years trend. However, we need a very long time series of data to confirm it. 
Apart from this periodicity there is a peak corresponding about $\approx$24 years with a confidence level $>$99.9 \%. Other peaks between 6,--11 years appear to represent the life of active period during different solar cycles and 11 year solar cycle period. The power spectra of average intensity of plages, central region of plages and EN are similar for larger periodicity $>$10 years but that of EN appear to be noisy for periods $<$10 years. The power spectrum plot for AN indicates a general periodic behavior of the small-scale activity with a period of $\approx$10.5 years. Very long period variation in the intensity of AN may be difficult to detect, because of small difference in the limit of minimum and maximum intensity of AN, even if it exists.

\subsection{Ca-K Index Considering Area and Intensity of Features}

\begin{figure}
\begin{center} 
\includegraphics[width=\textwidth]{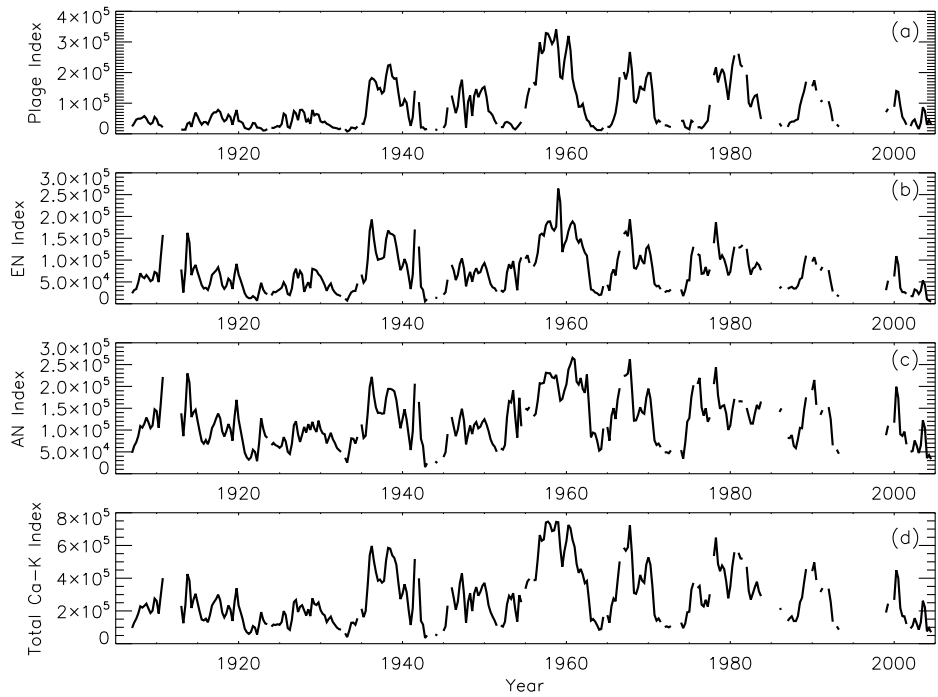} 
\end{center}
\caption{Panel a shows half yearly averaged KO plage index at an interval of three months for the period 1906,--2005. Panels b and c show the Ca-K index variations for the EN and AN network, respectively. Panel d indicates total Ca-K index considering the contribution from plages, enhanced, and active networks. The gaps represent no observational data available during that period.}
\label{fig:11}
\end{figure}

After determining the intensity of each pixel identified as the plage, we added the intensity of all these pixels having value of the intensity contrast $>$ 1.35 and area $>$ one~arcmin$^{2}$. We define the plage index (PI) for the day as, PI = PA$\times$IC. where, PA is the plage area on the solar surface expressed in arcsec$^{2}$ and IC is the intensity contrast, considering the area of each pixel as 0.74~arcsec$^{2}$. Similarly, we computed the EN and AN index on day to day basis. Top panel of Figure \ref{fig:11} shows the half yearly average plage index for the period of 1906 -- 2005. Middle two panels of this figure show the EN and AN index for the same period. The amplitude of plage index appears to be twice the amplitude of EN and AN indexes during the active phase of solar cycles. The temporal variations in  these three parameters appear to be correlated and they all are in phase. 
Though, the variations in the plage areas and index appear to dominate the solar-cycle 
variations, the contributions due to variations in the EN and AN indexes need to be considered in modeling the chromospheric solar cycle variations. This is because the combined network (EN and AN) index amplitude becomes comparable to the plage index. It may be noted that variation in the average plage intensity is large for different solar cycles as compared to variation in intensity of EN and AN for different solar cycles. Finally, we have  added all the three indexes to derive the total Ca-K index on day-to-day basis and plotted half yearly averaged total Ca-K index at an interval of three months in bottom most panel of Figure \ref{fig:11} for the period of 1906 -- 2005. This data will be useful for the study of solar cycle variations of chromosphere.

\section{Summary and Discussions}

We have digitized the Ca-K line spectroheliograms obtained at Kodaikanal observatory for the period of 1906 to 2005 with a  pixel resolution of 0.86 arcsec, using low noise CCD camera with 16-bit read out to study the long term variation of chromospheric features. We have developed a computer code to align, calibrate,  removal of limb darkening, stray light, and normalize the quiet chromospheric intensity to a uniform value of one. We have identified the plages, EN, AN, and QN which are by using threshold values of intensity contrast and area parameters determined empirically. The plage areas determined from the KO data agrees well with those derived from the MWO data for the period of 1906 -- 1985 and earlier measurements made by \citet{tlatov2009} for KO and NSO data for the period  of 1965 -- 2002. 

We have studied the variation in intensity contrast and the areas of the chromospheric features with the time and calculated the total Ca-K index for the year 1906 -- 2005. Uniformity of the time series and the precision analysis of the data using the code developed by us made it possible to study the long term variations in the intensity contrast of plages, EN and AN with an interval of six months using yearly averaged data. From the averaged data over a period of three years, the plage- intensity contrast appears to be about 1.85 during the active phases of Solar Cycles number 19 and 20, whereas it is about 1.55 during Solar Cycles 14 and 15. The intensity contrast for plages is around 1.50 for 22 and 23 Solar Cycles. The intensity of plages for other solar cycles has intermediate values indicating that intensity of plages vary with very long time period in addition to the variations with phase of solar cycle. These values imply that on an average plages are 20 \%, brighter in Solar Cycles 19 and 20 than during the active phase of Cycles 14, 15, 22, and 23. If we assume the intensity contrast as defined by 
\citet{worden1998} which means intensity of the quiet chromosphere is considered as zero then the average plage-intensity contrast during the Cycles 19 and 20 becomes about 50 \%, more than that during the Cycles 14, 15, 22, and 23.  Similarly, the intensity of EN during the active phase of Cycles 19 and 20 is more than the Cycles 14, 15, 22, and 23 by about 5 \%. Further, we found that yearly averaged intensity contrast of AN, representing small-scale activity over the whole of the solar surface varies by $\approx$1 \% with the phase of solar cycle, being larger at the maximum phase. 

It may be noted that \citet{tlatov2009} did not find any change in the plage intensity contrast from the data of KO, and NSO during their period of analysis. They found that the average plage intensity contrast for KO, MWO, and NSO is 1.33, 1.50, and 2.09, respectively. Using the same data set, \citet{bertello2010} show that the plage contrast varies with time and was maximum around 1957 -- 58 during the strongest Cycle 19. The analysis done by \citet{tlatov2009} indicated the intensity contrast of the plages during the maximum phase of Cycle 19 for the 
MWO data is more compared to other solar cycles. They reasoned that in Solar Cycles 18 and 19 many sunspots appeared on the disk.
The bandpass of the exit slit of the KKL is much larger than MWO. Because of 
this, the dark sunspots appear inside the plages in KKL data compared to MWO data. 
The analysis by \citet{tlatov2009} shows a significant increase in the full-disk plage contrast during the maximum phase of Cycle 19, as compared to other solar cycles. This was 
attributed to a combination of the masking effect of sunspots in calculating the plage index 
and in the MWO exit slit width being narrower during that period \citep{bertello2010}.
Further, \citet{penn2006} found that magnetic fields and temperature of the sunspots are less during the minimum phase compared to those at maximum phase.

Further, it may be noted that after removing the contribution from solar activity from the basal line profiles, \citet{pevtsov2013} find a weak dependency of intensity in the line core (K3) of basal profiles with the phase of the solar cycle. The weak dependency on the solar-cycle phase may be due to small-scale activity. We also find 11 year quasi-periodicity in the intensity of active network. We could find the variation in the intensity contrast with time because of high spatial resolution, better photometric accuracy of digitization, making correction for the intensity vignetting  in the images due to instrument in addition to limb darkening effect and careful normalization of the quiet chormosphere. \citet{nindos1998} and \citet{ortiz2005} 
have found that there is a strong correlation between areas of Ca-K emission in active regions and underlying photospheric magnetic field. We, therefore, conclude that our finding of long term variation in the intensity of plages and EN implies that on an average, strength of magnetic field in active regions is larger for stronger solar cycles represented by more number and area of sunspots during that cycle. In this article, we reported the long-time variations of plage intensity, longer than the solar cycle period for the first time. It needs to be investigated how this change in intensity of plage over the long-time period contributed to the TSI and changes in the temperature of the Earth.

A group of researchers at the Indian Institute of Astrophysics is in the process of calibrating the Ca-K images without using the step-wedge. This is achieved by using the quiet sun profile which is assumed to be constant during the period of 100 years 
of observations. The preliminary results show that there is not much difference in the results obtained from both methods (with step wedge and quiet sun profile). However, more quantitative analsysis need to be done.

\begin{acks}

\par We thank  T. G. Priya, K. Amareswari, Nazia, S.M.A.Fathima, S. Kamesh, R. Janani, M. Aysha Banu, and P. Sathish for their help in digitizing the photographic images and running of computer programs to analyse the data. We also thank large number of observers who have obtained the valuable data and maintained the data in good environment conditions.  This project to digitize the photographic solar data at KO was planned, designed, digitized, calibrated the data by Jagdev Singh and a team trained by him. The authors thank the referee for insightful comments on the original version of the manuscript.

\end{acks}

\section*{Disclosure of Potential Conflicts of interest}
The authors declare that they have no conflicts of interest.

\end{document}